\pdfoutput=1

\documentclass[11pt,twoside,a4paper,cmspaper,final,collab]{cms-tdr}
\def\svnVersion{460783P}\def\cmsCernNoTag{CERN-EP-2018-001}\def\cmsCernDate{\today}\def\cmsMessage{Published in the Journal of High Energy Physics as \href{http://dx.doi.org/10.1007/JHEP04(2018)073}{\doi{10.1007/JHEP04(2018)073.}}}

\begin{document}\cmsNoteHeader{EXO-16-058}

\hyphenation{had-ron-i-za-tion}
\hyphenation{cal-or-i-me-ter}
\hyphenation{de-vices}
\RCS$HeadURL: svn+ssh://svn.cern.ch/reps/tdr2/papers/EXO-16-058/trunk/EXO-16-058.tex $
\RCS$Id: EXO-16-058.tex 450506 2018-03-11 13:13:06Z swmukher $

\newlength\cmsTabSkip\setlength{\cmsTabSkip}{1ex}
\cmsNoteHeader{EXO-16-058}
\title{Search for lepton-flavor violating decays of heavy resonances and quantum black holes to $\Pe\Pgm$ final states in proton-proton collisions
at $\sqrt{s}=13\TeV$}

\date{\today}

\abstract{
A search is reported for heavy resonances decaying into $\Pe\Pgm$ final states in proton-proton collisions recorded by the CMS experiment
at the CERN LHC at $\sqrt{s}=13\TeV$, corresponding to an integrated luminosity of 35.9\fbinv. The search focuses on resonance
masses above 200\GeV. With no evidence found for physics beyond the standard model in the $\Pe\Pgm$ mass spectrum, upper limits are set
at 95\% confidence level on the product of the cross section and branching fraction for this lepton-flavor violating signal.
Based on these results, resonant {\Pgt} sneutrino production in R-parity violating supersymmetric models is excluded for
masses below 1.7\TeV, for couplings $\lambda_{132}=\lambda_{231}=\lambda'_{311}=0.01$. Heavy \PZpr gauge bosons with lepton-flavor violating
transitions are excluded for masses up to 4.4\TeV.
The $\Pe\Pgm$ mass spectrum is also interpreted in terms of non-resonant contributions from quantum black-hole
production in models with one to six extra spatial dimensions, and lower mass limits are found between 3.6 and 5.6\TeV.
In all interpretations used in this analysis, the  results of this search improve previous limits by about 1\TeV.
These limits correspond to the most sensitive values obtained at colliders.}

\hypersetup{
pdfauthor={CMS Collaboration},
pdftitle={Search for lepton-flavor violating decays of heavy resonances and quantum black holes to emu final states in proton-proton collisions at sqrt s = 13 \TeV},
pdfsubject={CMS},
pdfkeywords={CMS, physics, LFV, SUSY, QBH}}

\maketitle
\section{Introduction}
Several extensions of the standard model (SM) predict the existence of heavy particles that undergo lepton-flavor
violating (LFV) decays, thereby motivating searches for deviations from the SM in $\Pe\Pgm$ final states.
This paper reports a search for such phenomena in the $\Pe\Pgm$ invariant mass spectrum $m_{\Pe\Pgm}$.
The analysis is based on data corresponding to an integrated luminosity of 35.9\fbinv collected
in proton-proton ($\Pp\Pp$) collisions at $\sqrt{s}=13\TeV$ in the CMS detector at the CERN LHC. The search strategy is
designed to be model independent as much as possible. The results are interpreted
in terms of the characteristics of the following predicted states: a {\Pgt} sneutrino ($\PSgn_\Pgt$), which can be the
lightest supersymmetric particle (LSP)~\cite{Farrar:1978xj,deCampos:2008av,Bartl:2003fh} in R-parity violating (RPV)
supersymmetric (SUSY) models~\cite{Barbier:2004ez}, a heavy \PZpr gauge boson in LFV
models~\cite{Langacker:2008yv}, and quantum black holes (QBHs)~\cite{Calmet:2008dg,Meade:2007sz}. The theoretical underpinnings
in the context of this search are introduced below.

In RPV SUSY models, lepton flavor and lepton number are violated at the lowest Born level in interactions
between fermions and their superpartners, where the $\PSgn_\Pgt$ can be the LSP. For resonant $\PSgn_\Pgt$ signals,
the trilinear RPV part of the superpotential can be expressed as
\begin{equation*}
W_{\text{RPV}} = \frac{1}{2} \lambda_{ijk} L_i L_j \overline{E}_k
+ \lambda'_{ijk} L_i Q_j \overline{D}_k,
\end{equation*}
where: $i$, $j$, and $k$ are generation indices; $L$ and $Q$ are
the $\text{SU}(2)_L$ doublet superfields of the leptons and quarks; and $\overline{E}$ and $\overline{D}$ are the respective $\text{SU}(2)_L$
singlet superfields of the charged leptons and down-like quarks.

For simplicity, we suppose that all RPV couplings vanish, except for
$\lambda_{132}$, $\lambda_{231}$, and $\lambda'_{311}$, which are connected to the
production and decay of the  $\PSgn_\Pgt$, and we consider a SUSY mass hierarchy with $\PSgn_\Pgt$ as the
LSP. In this model, the $\PSgn_\Pgt$ can be produced resonantly in $\Pp\Pp$ collisions via the $\lambda'_{311}$ coupling,
and can decay either into $\Pe\Pgm$ via the $\lambda_{132}$ and $\lambda_{231}$ couplings, or into
$\cPqd\PAQd$ via the $\lambda'_{311}$ coupling. We consider only the $\Pe\Pgm$ final state,
and assume $\lambda_{132} = \lambda_{231}$. This analysis
considers only $\PSgn_\Pgt$ that decay promptly and not
long-lived $\PSgn_\Pgt$~\cite{Graham:2012th}, which could provide events with {\Pe} and {\Pgm} tracks from a displaced vertex.

An extension of the SM through the addition of an extra U(1) gauge symmetry provides a massive \PZpr vector boson~\cite{Langacker:2008yv}.
In our search, we assume that the \PZpr boson has couplings similar to the Z boson in the SM, but that the \PZpr boson can also decay to the LFV {\Pe}{\Pgm}
final state with a branching fraction of 10\%. The resulting \PZpr width is approximately 3\% of its mass for masses above the \ttbar threshold.

\begin{figure}[h]
\centering
\includegraphics[width=0.33\textwidth]{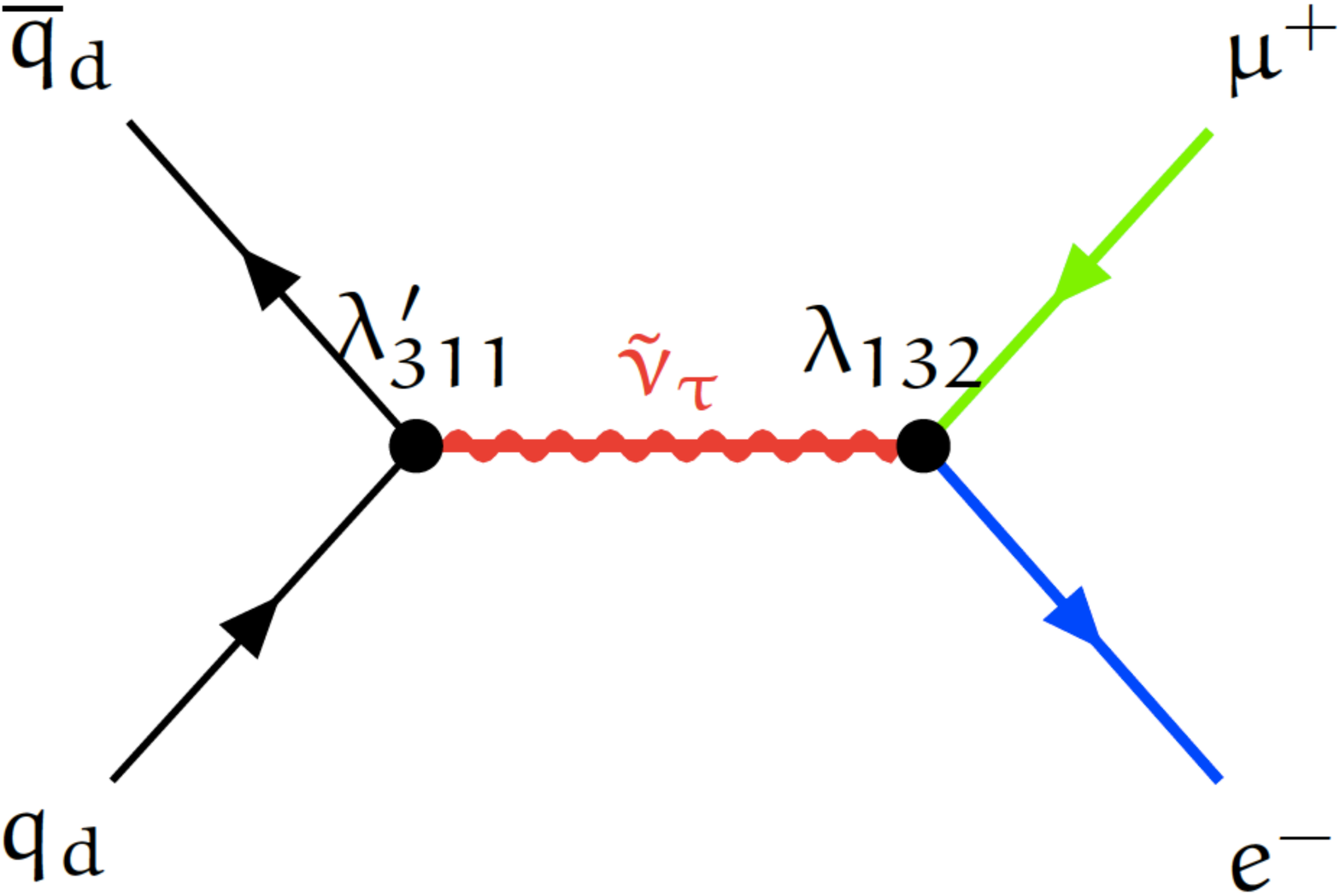}~
\includegraphics[width=0.33\textwidth]{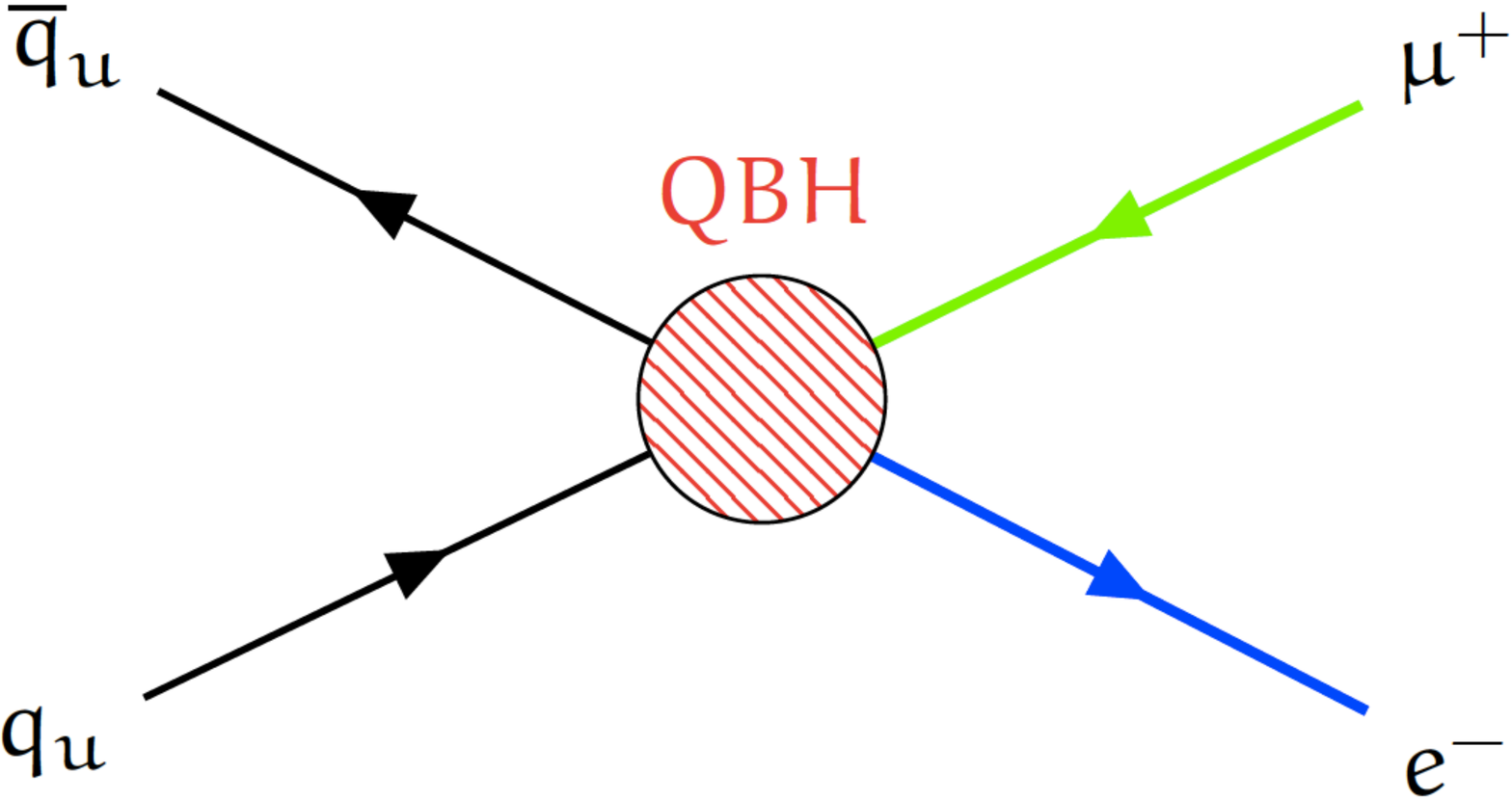}~
\includegraphics[width=0.33\textwidth]{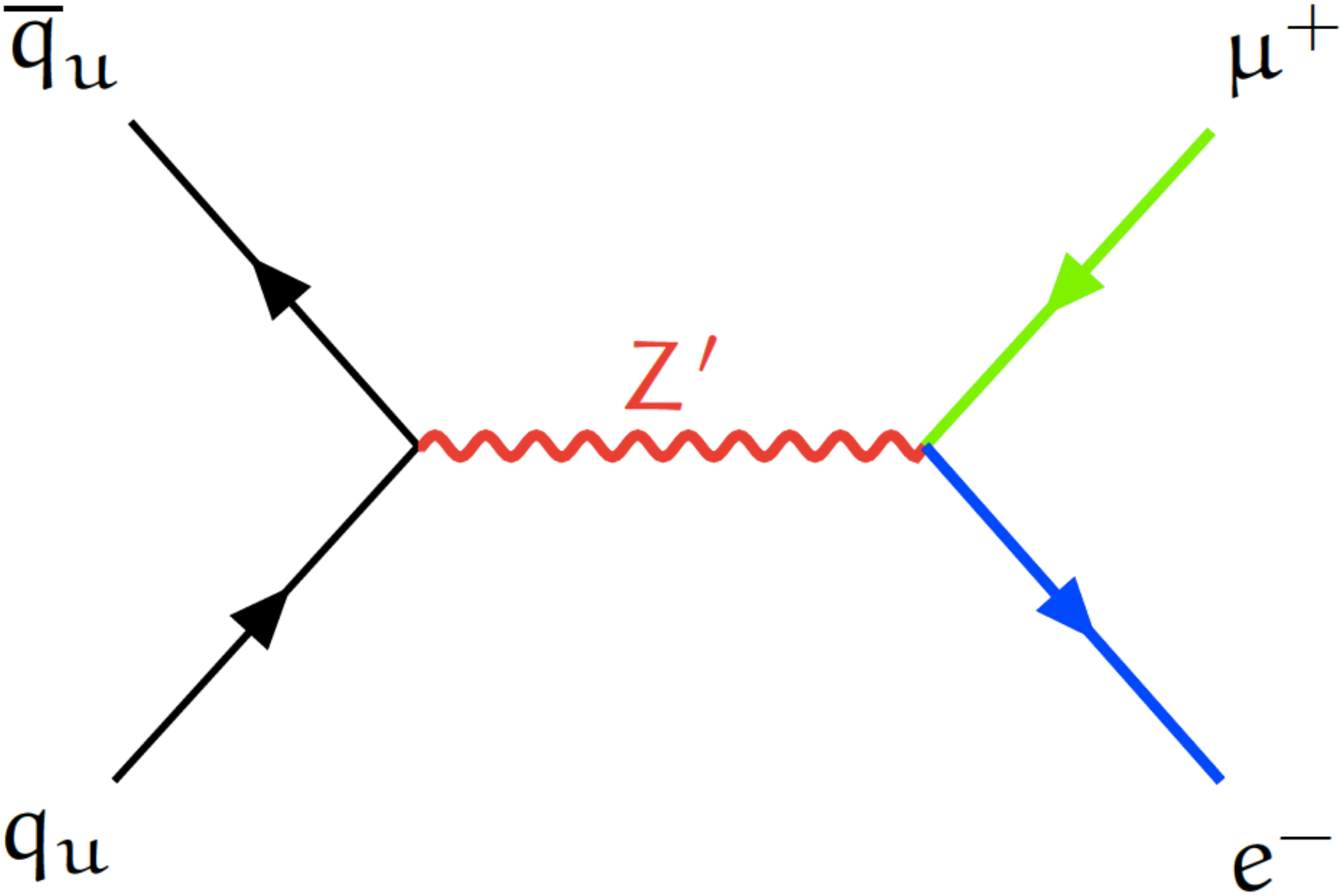}
\captionof{figure}[]{Leading order Feynman diagrams considered in our search. Left: Resonant production
of a {\Pgt} sneutrino in an RPV SUSY model that includes the subsequent decay into an electron and a muon. The $\PSgn_\Pgt$ is produced from the
annihilation of two down quarks via the $\lambda'_{311}$ coupling, and then decays via the $\lambda_{132} = \lambda_{231}$ couplings into the
electron muon final state. Middle: Production of quantum black holes in a model with
extra dimensions that involves subsequent decay into an electron and a muon. Right: Resonant production
of a \PZpr boson with subsequent decay into an electron and a muon.}
\label{fig:feynman}
\end{figure}

Theories that invoke extra spatial dimensions can offer effective fundamental Planck scales in the TeV region. Such theories also provide the
possibility of producing microscopic black holes~\cite{Calmet:2008dg,Meade:2007sz} at the LHC. In contrast to semiclassical thermal
black holes that can decay to high-multiplicity final states, QBHs are nonthermal objects, expected to decay predominantly to
pairs of particles. We consider the production of spin-0, colorless, neutral QBHs in a model with LFV~\cite{Gingrich:2009hj},
in which the cross section for QBH production depends on the threshold mass $m_{\text{th}}$ in $n$ additional
spatial dimensions. The $n = 1$ possibility corresponds to the Randall--Sundrum (RS) brane-world model~\cite{Randall:1999ee}, and $n>1$
corresponds to the Arkani--Hamed--Dimopoulos--Dvali (ADD) model~\cite{ArkaniHamed:1998rs}. While the resonant $\PSgn_\Pgt$ and
\PZpr signals generate narrow peaks in the invariant mass spectrum of the $\Pe\Pgm$ pair, the distribution of the QBH signal is
characterized by a sharp edge at the threshold of QBH production, followed by a monotonic decrease at larger masses.
Feynman diagrams for all these three models are shown in Fig.~\ref{fig:feynman}.

Similar searches in the $\Pe\Pgm$ mass spectrum have been carried out by the CDF~\cite{Aaltonen:2010fv} and D0~\cite{Abazov:2010km}
experiments at the Fermilab Tevatron in $\Pp\Pap$ collisions at a center-of-mass energy of 1.96\TeV and by the ATLAS
and CMS experiments at the LHC in $\Pp\Pp$ collisions at center-of-mass energies of 8\TeV~\cite{Aad:2015pfa,Khachatryan:2016ovq} and
13\TeV~\cite{TheATLAScollaboration:2015ucd}. The search by CMS at 8\TeV has an integrated luminosity of 19.7\fbinv, and excludes
$\PSgn_\Pgt$ masses up to 1.28\TeV for $\lambda_{132} = \lambda_{231} = \lambda'_{311} = 0.01$. The search performed by ATLAS at 13\TeV
with 3.2\fbinv of luminosity excludes \PZpr bosons with mass up to $m_{\PZpr} = 3.01\TeV$. The present search significantly extends these limits.

\section{The CMS detector}
The central feature of the CMS apparatus is a superconducting solenoid of 6\unit{m} internal diameter, providing a
magnetic field of 3.8\unit{T}. A silicon pixel and strip tracker, a lead tungstate crystal electromagnetic calorimeter
(ECAL), and a brass and scintillator hadron calorimeter (HCAL), each composed of a barrel and two end sections, reside
within the solenoid volume. Forward calorimeters extend the pseudorapidity ($\eta$) coverage provided by the barrel
and end calorimeters. Muons are detected in gas-ionization chambers embedded in the steel flux-return yoke outside
the solenoid. A more detailed description of the CMS detector, together with a definition of the coordinate system
and kinematic variables, can be found in Ref.~\cite{Chatrchyan:2008zzk}.

\section{Event selection}
The search is designed to be inclusive and model independent, requiring
at least one prompt, isolated electron
and at least one prompt, isolated muon in the event. This minimal selection also facilitates a reinterpretation of
the results in terms of models with more complex signal topologies than the single $\Pe\Pgm$ pair.
Events that satisfy single-muon and single-photon triggers~\cite{Khachatryan:2016bia} with respective transverse
momentum (\pt) thresholds of 50 and 175\GeV for muons and photons are selected for analysis. Electromagnetic
energy deposited by an electron in the calorimeter activates the photon trigger used to record our events. The
photon trigger is therefore as efficient as the corresponding electron trigger, while its weaker isolation
requirements yield an event sample that can also be used in sideband analyses to estimate the background to the signal.

Electrons and muons are reconstructed and identified using standard CMS algorithms, described in
Refs.~\cite{Khachatryan:2015hwa,Chatrchyan:2012xi}.

To reconstruct an electron candidate, energy depositions in the ECAL are  first combined into clusters, assuming
that each  cluster represents a single particle. The clusters are then combined in a way consistent with bremsstrahlung
emission, to produce a single ``supercluster'', which represents the electron or photon. These superclusters are
used to seed tracking algorithms, and if a resulting track is found, it is associated to the supercluster to
form an electron candidate. The electron candidate must pass the high-energy electron pairs (HEEP)
selection~\cite{Khachatryan:2015hwa}, which requires the energy deposition in the ECAL to be consistent with that of
an electron. The sum of the energy in the HCAL within a cone of $\Delta R = 0.15$ centered around the electron candidate,
must be less than 5\% of its energy, after it is corrected for jet activity unrelated to the electron. The
electron candidate must have a well-matched, prompt track in the $\eta$-$\phi$ plane that has no more than one hit
missing in the inner portion of the tracker. The HEEP selection also requires electrons to be isolated, the requirement
for which is that the scalar-\pt sum of tracks within a cone of radius $\Delta R = 0.3$ around the candidate direction, excluding the candidate's track,
is less than 5\GeV, and the \pt sum of energy depositions in the calorimeters within this cone, taking account of small $\eta$-dependent
offsets, is less than 3\% of the \pt of the candidate.

To reconstruct a muon candidate, hits are first fitted
separately to trajectories in the inner-tracker detector, and in the outer-muon system. The two trajectories are then combined in a
global-muon track hypothesis.
Muon candidates are required to have $\pt>53\GeV$ and to
fall into the acceptance region of $\abs{\eta}<2.4$. The transverse and longitudinal impact parameters of muon candidates
relative to the primary vertex must be less than 0.2\unit{cm} and $<$0.5\unit{cm}, respectively.
The reconstructed vertex with the largest value of summed physics-object $\pt^2$ is taken to be the primary $\Pp\Pp$ interaction vertex.
The physics objects chosen are those that have been defined using information from the various subdetectors,
including jets, charged leptons, and the associated missing transverse momentum,
which is defined as the negative vector sum of the \pt of those jets and charged leptons, measured in the silicon tracker.
The track of the muon candidate must have at least one hit
in the pixel detector and hits in at least six silicon-strip layers, and must contain
matched segments in at least two muon detector planes. To suppress backgrounds arising from muons
within jets, the scalar-\pt sum of all other tracks in the tracker within a cone of
$\Delta R = \sqrt{\smash[b]{(\Delta\eta)^2+(\Delta\phi)^2}} = 0.3$ around the muon candidate track,
where $\eta$ and $\phi$ are the pseudorapidity and azimuth angle of a track, is required to have
less than 10\% of the \pt of the muon candidate. The relative uncertainty
in \pt of the muon track is required to be smaller than 30\%.

To reduce loss in signal efficiency from misidentification of the sign of the electron's or muon's charge at large \pt, the
electron and muon are not required to have opposite charges. Since highly energetic muons can produce bremsstrahlung in the ECAL along the
direction of the inner-muon trajectory, such muons can be misidentified as electrons. An electron candidate is therefore
rejected if there is a muon candidate with \pt greater than 5\GeV whose track has $\Delta R<0.1$ relative to the electron candidate's track.
Only one $\Pe\Pgm$ pair is considered per event. When there is more than one {\Pe}{\Pgm} candidate, the pair with the highest
invariant mass is selected for analysis.

The statistical interpretation is done based on the shape of the invariant $\Pe\Pgm$ mass distribution of the signal
as well as the background.

\section{Signal simulation} \label{Sig_Sim}
The RPV SUSY $\PSgn_\Pgt$, \PZpr, and QBH signal events are generated at leading order (LO) precision, using
the \CALCHEP~3.6~\cite{Belyaev:2012qa}, \PYTHIA~8.203~\cite{Sjostrand:2007gs}, and \textsc{qbh}~2.0~\cite{Gingrich:2009da} Monte Carlo (MC) generators,
respectively. The relative width of the \PZpr signal is taken as
3\% of its mass, and interference between the SM Z and \PZpr bosons is ignored. All simulated signal events use \PYTHIA for
hadronization and CUETP8M1 provides the underlying-event tune~\cite{Khachatryan:2015pea}. The RPV and QBH signals
are generated with the CTEQ6L~\cite{Pumplin:2002vw} parton distribution functions (PDF) while the \PZpr boson signals are simulated using the
NNPDF 3.0 PDF sets~\cite{Ball:2014uwa}. The LO RPV SUSY $\PSgn_\Pgt$ signal event yield is normalized to a next-to-leading order (NLO)
calculation of the production cross section; in this calculation the factorization and renormalization scales are set to the mass of
the $\PSgn_\Pgt$. The generated events are processed through a full simulation of
the CMS detector, based on \GEANTfour~\cite{Agostinelli:2002hh,Allison:2006ve,ALLISON2016186}. The simulated events
incorporate additional pp interactions within the same or a nearby bunch crossings, termed pileup, that are weighted
to match the measured distribution of the number of interactions per bunch crossing in data. The simulated event samples
are normalized to the integrated luminosity of the data. The products of the total acceptance and
efficiency for the three signal models in this analysis are determined through MC simulation. The trigger and
object reconstruction efficiencies are corrected to the values measured in data.
The selection efficiencies for the RPV $\PSgn_\Pgt$, \PZpr and QBH signals are $\approx$60\%, 60\%, and 55\% when the resonance mass or mass
threshold is 1\TeV and $\approx$66\%, 64\%, and 63\% when the resonance mass or mass threshold is 4\TeV, respectively.

\section{Background estimation}
The SM backgrounds contributing to the $\Pe\Pgm$ final state are divided into two categories. The
first category comprises events with at least two real, isolated leptons; while the second category comprises
events that include either jets or photons, misidentified as isolated leptons, or jets with leptons
from heavy-flavor decays, both of which we refer to as fake background.

The expected SM background from processes with two real leptons is obtained from MC simulation. This background consists
mostly of events from \ttbar or {\PW}{\PW} production; the former process is dominant at lower masses
and the latter becomes equally important above $m_{\Pe\Pgm} \approx 1\TeV$. Other real lepton backgrounds estimated from
MC simulation involve diboson contributions from {\PW}{\cPZ} and {\cPZ}{\cPZ} events, single top quark production,
and Drell--Yan (DY)
production (\ie $\Pq\Paq \to $ virtual \cPZ/$\Pgg$ $\to$ two leptons of opposite
charge) in the {\Pgt}{\Pgt} channel. The $\text{DY}\to\Pgt\Pgt$
background is generated at NLO using the \MGvATNLO~2.2.2~\cite{Frederix:2012ps,Alwall:2014hca} event
generator. The cross sections used to normalize the contribution of these backgrounds
are calculated at next-to-next-to-leading order for
WW~\cite{Gehrmann:2014fva}, ZZ~\cite{Cascioli:2014yka}, single top quark~\cite{Berger:2016oht}, and \ttbar~\cite{Czakon:2013goa}
processes, and also at NLO accuracy for {\PW}{\cPZ}~\cite{Campanario:2010hp} and DY~\cite{Andonov:2009nn} events. All background
processes are simulated using the NNPDF 3.0 PDF. For all background simulations, \PYTHIA is used for hadronization
and CUETP8M1 as the underlying event tune.

The main sources of fake background in the $\Pe\Pgm$ selection  are from {\PW}+jets, {\PW}$\Pgg$ and DY+jets production,
where a jet or a photon is misidentified as an electron or a muon. The {\PW}$\Pgg$ process also contributes to the prompt
background category through the internal conversion of $\Pgg$ to leptons. The QCD-multijet process provides
subleading contributions to the fake lepton background. An estimate of the {\PW}$\Pgg$ background is
obtained at LO from MC simulation based on the \MGvATNLO event generator.
DY+jets background in {\Pe}{\Pe} and {\Pgm}{\Pgm} channels are generated
at NLO using the \POWHEG~2.0~\cite{Frixione:2007vw,Nason:2004rx,Alioli:2010xd} event generator.
A background estimate
based on a control sample in data is made using jet-to-electron misidentification rates ($F$) to determine the
$m_{\Pe\Pgm}$ contributions from {\PW}+jets and multijet distributions. The jet-to-electron misidentification
rate is measured in data, using a control sample collected with a single electromagnetic-cluster trigger. Data
sidebands are used to evaluate the contributions to the control sample from genuine electrons and from photons
misidentified as electrons. The jet-to-electron misidentification rate is then defined as the number of jets
passing the full electron selection divided by the number of jet candidates in the sample. The rate is quantified in
bins of \pt and $\eta$. The measured rate is used to estimate the {\PW}+jets and multijet contributions using data
containing muons that pass the single-muon trigger and the full muon selection, and the number of electron candidates satisfying
relaxed selection requirements, but failing the full electron selection. Each event is weighted by the factor $F/(1-F)$
to determine the overall contribution from the jet backgrounds.  Contributions from processes other than {\PW}+jets and
multijet sources are subtracted from the sample, after correcting for the contribution from false events
to avoid double counting, which is done using MC simulated background events. Background from jets mimicking muons is estimated to be only
1\% of the total background, and is ignored in the analysis.

\section{Systematic uncertainties}
The uncertainty in the modeling of the
$\Pe\Pgm$ invariant mass distribution reflects the input of three types of systematic effects.

The first type includes those that affect the shape of the invariant mass
distribution, with the dominant uncertainty arising from
the leading \ttbar and subleading {\PW}{\PW} backgrounds. The \ttbar background
provides an uncertainty of $<$30\% in the total background yield at $m_{\Pe\Pgm} \approx 1\TeV$, which reduces to $<$10\% 
at $m_{\Pe\Pgm} \approx 2\TeV$ because of the reduced contribution of the \ttbar process to the total background yield.
The uncertainty in the {\PW}{\PW} background is estimated to be $\approx$2.5\% at $m_{\Pe\Pgm} \approx 1\TeV$.
This is estimated from the envelope of the resummed next-to-next-to-leading-logarithm calculation of the soft-gluon
contributions to the cross section at NLO, as presented in Ref.~\cite{Pecjak:2016nee}, using changes by factors of
2 and 0.5 implemented in the renormalization and factorization scales, respectively. Other uncertainties in the form of the invariant mass
distribution are due to the uncertainty in the muon momentum scale, which depends on the $\eta$ and $\phi$ of muons,
and leads to an uncertainty in the total background yield of $\approx$1.1\%, at $m_{\Pe\Pgm} = 500\GeV$, and
$\approx$25\% at $m_{\Pe\Pgm} = 2\TeV$. Uncertainty in the muon-efficiency scale factor is 2--3\% over the
whole mass range. Apart from that, a momentum-dependent, one-sided downward systematic uncertainty is applied to the muon
reconstruction and identification efficiency, to account for potential differences between simulated samples and data, in the response of the muon system
to muons that interact radiatively with the detector material.
This uncertainty is --1.6\% in the region $\abs{\eta}<1.6$, and --14.4\% in the region $1.6<\abs{\eta}<2.4$, for muons with momentum of 4\TeV.
Uncertainties in the electron \pt scale and resolution, the muon \pt resolution, and the pileup rate have negligible impact on
the total background. Uncertainty in the electron-efficiency scale factor is 2--3\% over the whole mass range. The
uncertainty associated with the choice of the PDF in the simulation is evaluated according to the PDF4LHC
prescription~\cite{Butterworth:2015oua,Ball:2012cx}.

Uncertainties of the second type
directly influence the normalization of the invariant mass distribution.
A systematic uncertainty of 2.5\%~\cite{CMS:2017sdi}
in the integrated luminosity is taken for the backgrounds and signals.
Among the uncertainties in the cross sections used for the normalization of various simulated backgrounds, the 5\%
uncertainty in the dominant \ttbar background is most significant. A systematic uncertainty of
50\% is applied to the estimate of jet background derived from data.

Uncertainties of the third type are associated with limited sizes of event samples in the MC
simulation of background processes. In contrast to all other uncertainties, they are not correlated between bins of the invariant mass distribution.

Taking all systematic uncertainties into
account, the resulting uncertainty in the background ranges from 15\% at $m_{\Pe\Pgm} = 200\GeV$ to
40\% at $m_{\Pe\Pgm} = 2\TeV$. The increase in systematic uncertainty with increasing mass does not affect
the sensitivity at large mass values, where the expected number of events from SM processes
becomes negligible. All these uncertainties are also considered in the estimation of theoretical signals.

\section{Results}
\begin{figure}[htp]
\centering
\includegraphics[width=.6\textwidth]{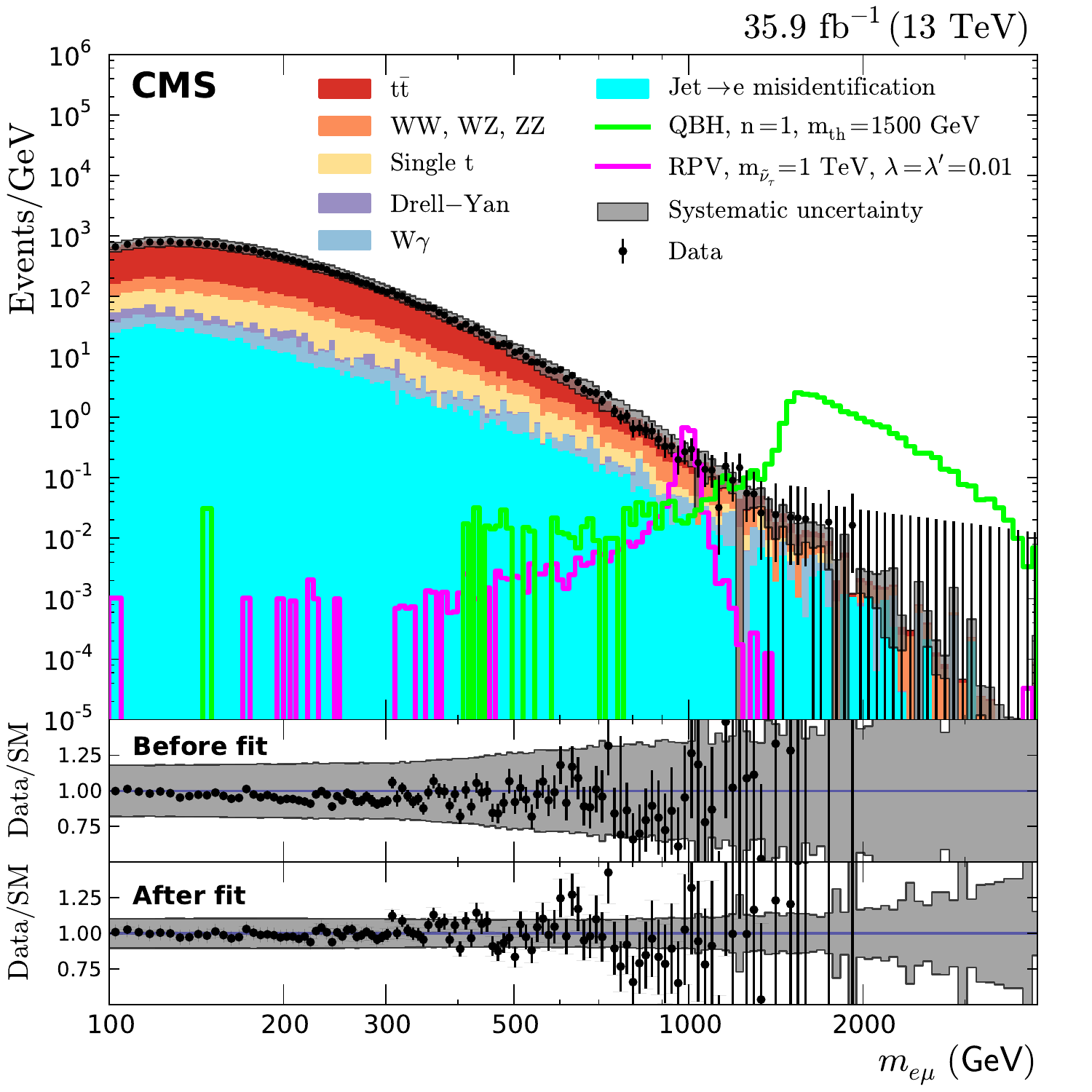}\\
\includegraphics[width=.6\textwidth]{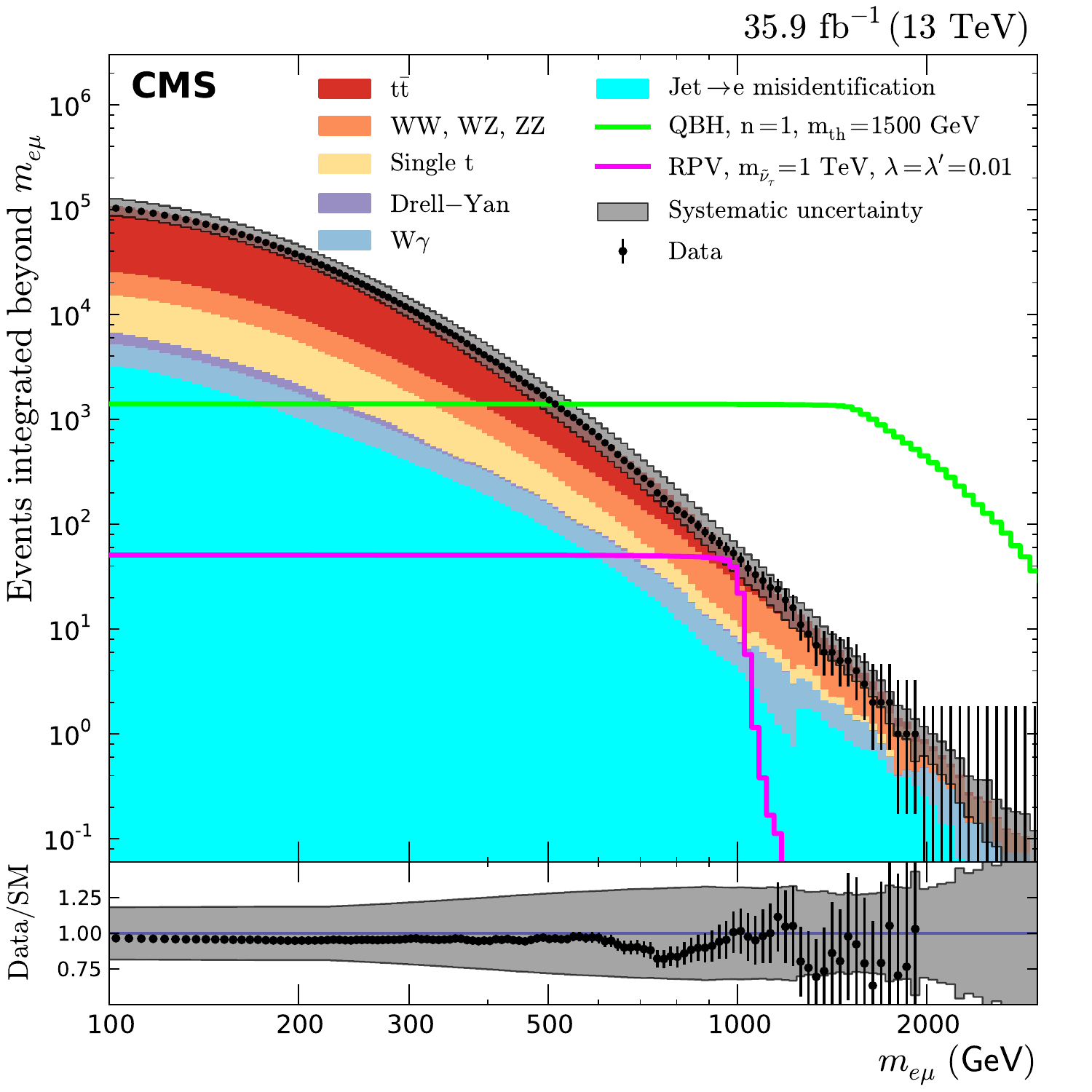}
\caption{Upper: The invariant mass distribution for selected $\Pe\Pgm$ pairs in data (black points
         with error bars), and stacked histograms representing expectations from SM processes before the fit. Also shown are the
         expectations for two possible signals. The two lower panels show the ratio of data to background
         expectations before and after the fit. The total systematic uncertainties are given by the gray bands.
         Lower: The cumulative (integral) distribution in events integrated beyond the chosen
         $m_{\Pe\Pgm}$. The lower panel shows the ratio of data to background predictions before the fit.
         Some events in the invariant mass distribution can have a negative event weight and result in a rise
         of the cumulative mass distribution. In both figures the label $\lambda$ refers to $\lambda_{132} = \lambda_{231}$,
         while $\lambda'$ stands for $\lambda'_{311}$.
}
\label{fig:emumass}
\end{figure}

The $\Pe\Pgm$ mass distribution for the selected events is shown in Fig.~\ref{fig:emumass}, together with
the corresponding cumulative distribution, integrated from the chosen $m_{\Pe\Pgm}$ threshold
to 10\TeV. The binning has been chosen to match the experimental resolution.
A version of the invariant mass distribution with a coarser binning, which reduces statistical fluctuations
and thus enhances the display of the simulated signals, is provided in the Appendix as Fig.~\ref{fig:emumass_simple}. A
simultaneous maximum likelihood fit to the data, including all systematic uncertainties, is performed.
Its effect may be seen by comparing the ratio of data to expected background, before and after the fit.
The fit represents a close estimate
of the distribution used in setting the upper limits on cross sections, as described below.

After selection, 4 events are observed in data in the region $m_{\Pe\Pgm}>1.5\TeV$, where the expectation
from SM backgrounds is $4.64 \pm 0.42\stat\pm 1.28\syst$. A detailed table showing the contribution from the
different background processes as well as the observed data is given in Tab.~\ref{tab:background composition}. The
largest observed value of $m_{\Pe\Pgm}$ is 1.89\TeV, and no significant excess is observed relative to
the SM expectation. Below, we set limits on the product of the signal cross section ($\sigma$) and the branching
fraction ($\mathcal{B}$) of signal to $\Pe\Pgm$.

\begin{table}[h]
\centering
\topcaption{Numbers of events for background processes, total background with its associated systematic uncertainties, and data, in four bins
         of $\Pe\Pgm$ invariant mass.
}
\begin{tabular}{lcccc}
\hline
Mass range (\GeVns)  & $m_{\Pe\Pgm} < 500$ & $500 < m_{\Pe\Pgm} < 1000$ & $1000 < m_{\Pe\Pgm} < 1500$ & $m_{\Pe\Pgm} > 1500$ \\ \hline
Jet$\to${\Pe} misidentification & 3601           & 82.8                  & 2.92                   & 0.849 \\
\PW$\Pgg$                             & 2462           & 56.2                  & 2.76                   & 0.562 \\
Drell--Yan                              & 2638           & 5.31                  & 0.343                  & 0.0145 \\
Single t                           & 9930           & 141                   & 2.81                   & 0.178 \\
\PW\PW, \PW\cPZ, \cPZ\cPZ                     & 11126          & 239                   & 13.0                   &  2.03 \\
\ttbar                               & 96754          & 971                   & 18.5                   & 1.01 \\ [\cmsTabSkip]
Total background                               & 126513         & 1495                  & 40.3                   & 4.64 \\
Systematic uncertainty                         & 23495          & 420                   & 13.5                   & 1.28 \\ [\cmsTabSkip]
Data                                           & 123150         & 1426                  & 41                     & 4 \\ \hline
\end{tabular}
\label{tab:background composition}
\end{table}

In the context of RPV SUSY model, the $\PSgn_\Pgt$ can be produced at the LHC through an $s$-channel $\Pq\Paq$
interaction, which gives rise to a narrow resonance signal. For coupling values considered in
this analysis, the intrinsic width of this signal is small compared to the detector resolution. To describe the response of the electromagnetic
calorimeter, a Crystal Ball
function~\cite{Oreglia:1980cs} is used to model the signal.
For each resonance mass,
two parameters, the product of signal acceptance and efficiency ($A \epsilon$) and the mass resolution,
define the line shape used for setting the limit. Both parameters are obtained from fits to MC signal events.
The parametric forms of $A \epsilon$ and mass resolution as a function of signal mass
are given in Table~\ref{tab:table1}, for the values discussed in Section~\ref{Sig_Sim}, and shown in the Appendix as
Figs.~\ref{fig:emu_eff} and \ref{fig:emu_res}, respectively.
The invariant mass resolution ranges from 2.2\%, at a resonance mass of 200\GeV to 3.1\% at a mass of 3\TeV.
The parametrization of the resonant line shape provides sensitivity for invariant masses between the
simulated signal points. The QBH signal exhibits a broader distribution with
a cliff at the threshold mass $m_{\text{th}}$ that is smeared out by detector resolution, and a tail at
larger masses that falls off because of the form of the proton PDF. The QBH signals are not parametrized but obtained from
the simulated invariant mass distributions. The \PZpr signals give rise to resonance forms that are also taken directly
from simulation.

\begin{table}[h]
\centering
\topcaption{Parametrization functions for the product of the acceptance and efficiency, and for the invariant mass resolution for the RPV signal.
         The value of $m_{\PSgn_\Pgt}$ is given in units of \GeV. The functions are shown in Figs.~\ref{fig:emu_eff} and \ref{fig:emu_res}
         in the Appendix.
}
\renewcommand{\arraystretch}{1.4}
\begin{tabular}{cc}
\hline
Parameter           &  Functional form \\  \hline 
$A \epsilon$  &  $ -0.838 + \frac{1.67\times 10^{-2}}{(m_{\PSgn_\Pgt}^{-1.02} + 1.0\times 10^{-2})} - 1.54\times 10^{-5} m_{\PSgn_\Pgt} $   \\ [\cmsTabSkip] 
Mass resolution     &  $ 1.79\times 10^{-2} + 1.47\times 10^{-5} m_{\PSgn_\Pgt} -3.87\times 10^{-9} m_{\PSgn_\Pgt}^2 + 4.34 \times 10^{-13} m_{\PSgn_\Pgt}^3  $  \\ \hline
\end{tabular}
\label{tab:table1}
\end{table}

An upper limit at 95\% confidence level (CL) on $\sigma \mathcal{B}$ is determined using a Bayesian binned-likelihood
approach~\cite{ATLAS:2011tau}, assuming a uniform prior for the signal cross section. The signal and background $m_{\Pe\Pgm}$ distributions
enter the likelihood binned to the nearest 1\GeV, which is well below the invariant mass resolution for masses $>$200\GeV. The
nuisance parameters associated with the systematic uncertainties are modeled through log-normal distributions for uncertainties
in the normalization. Uncertainties in the shape of the distributions are modeled through ``template morphing'' techniques~\cite{Baak:2014fta}.
A Markov Chain MC method~\cite{Allanach:2007qj} is used for
integration. After integrating over all nuisance parameters for each mass hypothesis, the posterior probability density is
calculated as a function of $\sigma \mathcal{B}$ for yields at 95\% CL upper limit.

\begin{figure}[htp]
\centering
\includegraphics[width=.64\textwidth]{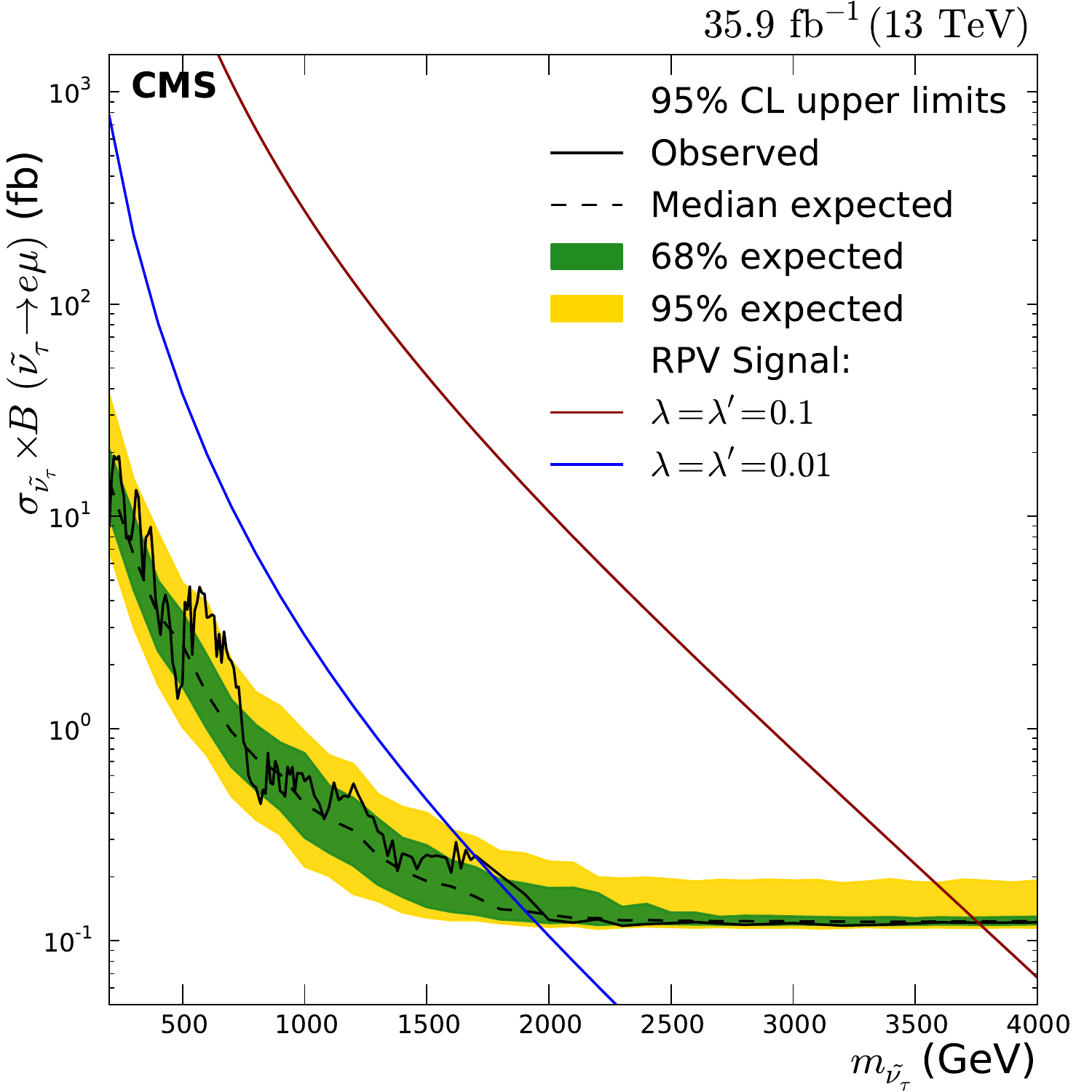}\\ \vspace{.1cm}
\includegraphics[width=.64\textwidth]{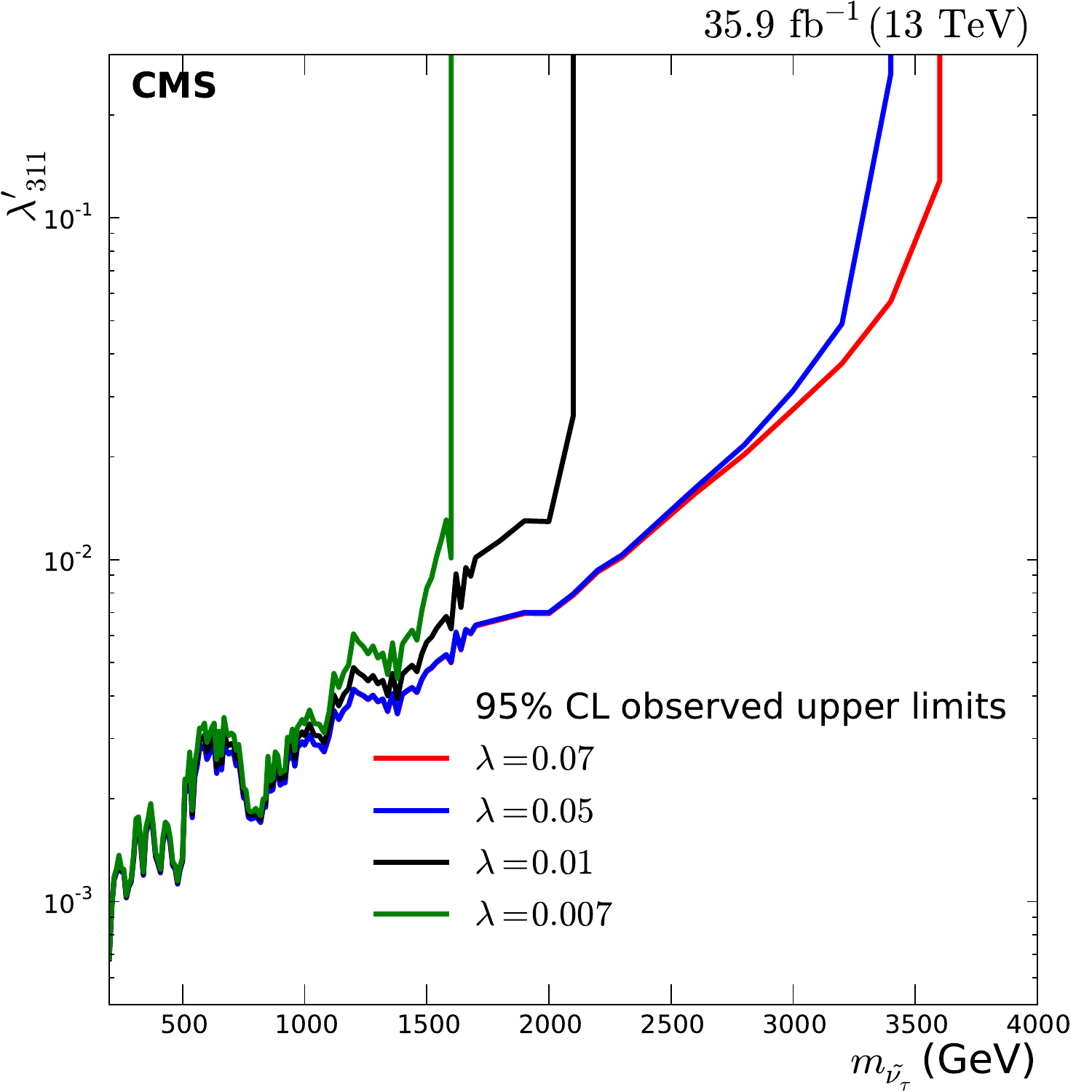}
\caption{Upper: Upper limits at 95\% CL on the product of the signal cross section and branching fraction for the $\PSgn_\Pgt$
         signal, as a function of the mass of the RPV resonance. The 68 and 95\% CL intervals on the median expected limits are
         indicated, respectively, by the inner green and outer yellow shadings. Predictions for an RPV SUSY model are shown for two values of
         the coupling parameter.
         Lower: Upper limits at 95\% CL on the RPV $\PSgn_\Pgt$ signal in the $(m_{\PSgn_{\Pgt}},\lambda'_{311})$
         parameter plane, for four values of $\lambda$, where the regions to the left of and above the limits are excluded.
         In both figures $\lambda$ refers to $\lambda_{132} = \lambda_{231}$, while $\lambda'$ stands
         for $\lambda'_{311}$.
}
\label{fig:RPVlimit}
\end{figure}

The limits at 95\% CL on $\sigma \mathcal{B}$ for the RPV $\PSgn_\Pgt$ signal are shown in Fig.~\ref{fig:RPVlimit}
(left). The signal cross section is calculated at NLO for the RPV couplings $\lambda_{132} = \lambda_{231} = \lambda'_{311} = 0.01$
and 0.1. The factorization and renormalization scales that enter the calculation are set to the mass of the $\PSgn_\Pgt$.
For RPV coupling $\lambda_{132} = \lambda_{231} = \lambda'_{311} = 0.01$, we obtain a lower mass limit of 1.7\TeV, while a limit of 1.9\TeV
is expected. For RPV coupling 0.1, we both observe and expect a 3.8\TeV
mass limit. In the narrow-width approximation, $\sigma \mathcal{B}$ scales with the RPV couplings
as~\cite{Khachatryan:2016ovq}:
\begin{equation*}
\sigma \mathcal{B} \propto (\lambda'_{311})^2[(\lambda_{132})^2 + (\lambda_{231})^2] / [3 (\lambda'_{311})^2 +  (\lambda_{132})^2 + (\lambda_{231})^2 ].
\end{equation*}
Using this relation and the observed upper bounds, we obtain limit contours in the $(m_{\PSgn_{\Pgt}},\lambda'_{311})$
parameter plane for several fixed values of $\lambda$. The result is shown in Fig.~\ref{fig:RPVlimit} (right).

\begin{figure}[htp]
\centering
\includegraphics[width=.65\textwidth]{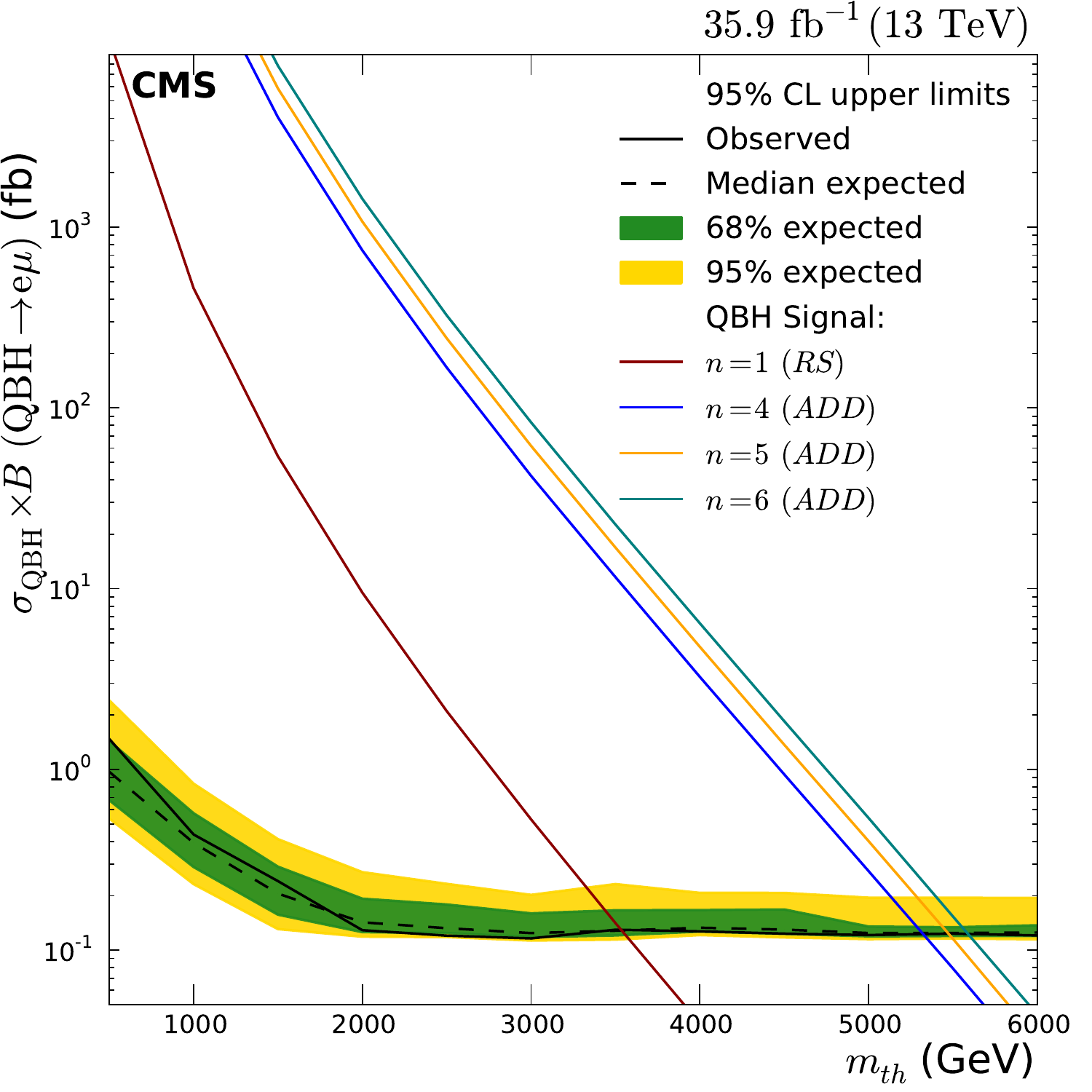}
\caption{Upper limits at 95\% CL on the median product of the signal cross section and the branching fraction
         for the QBH decay to $\Pe\Pgm$ as a function of threshold mass $m_\text{th}$. The 68 and 95\% CL
         intervals on the median are indicated, respectively, by the inner green and outer yellow shadings. Predictions are
         also shown for several models with large extra spatial dimensions, specifically for 1 extra dimension
         (RS) and for 4, 5, and 6 extra dimensions (ADD).
}
\label{fig:QBHlimit}
\end{figure}

\begin{figure}[htp]
\centering
\includegraphics[width=.65\textwidth]{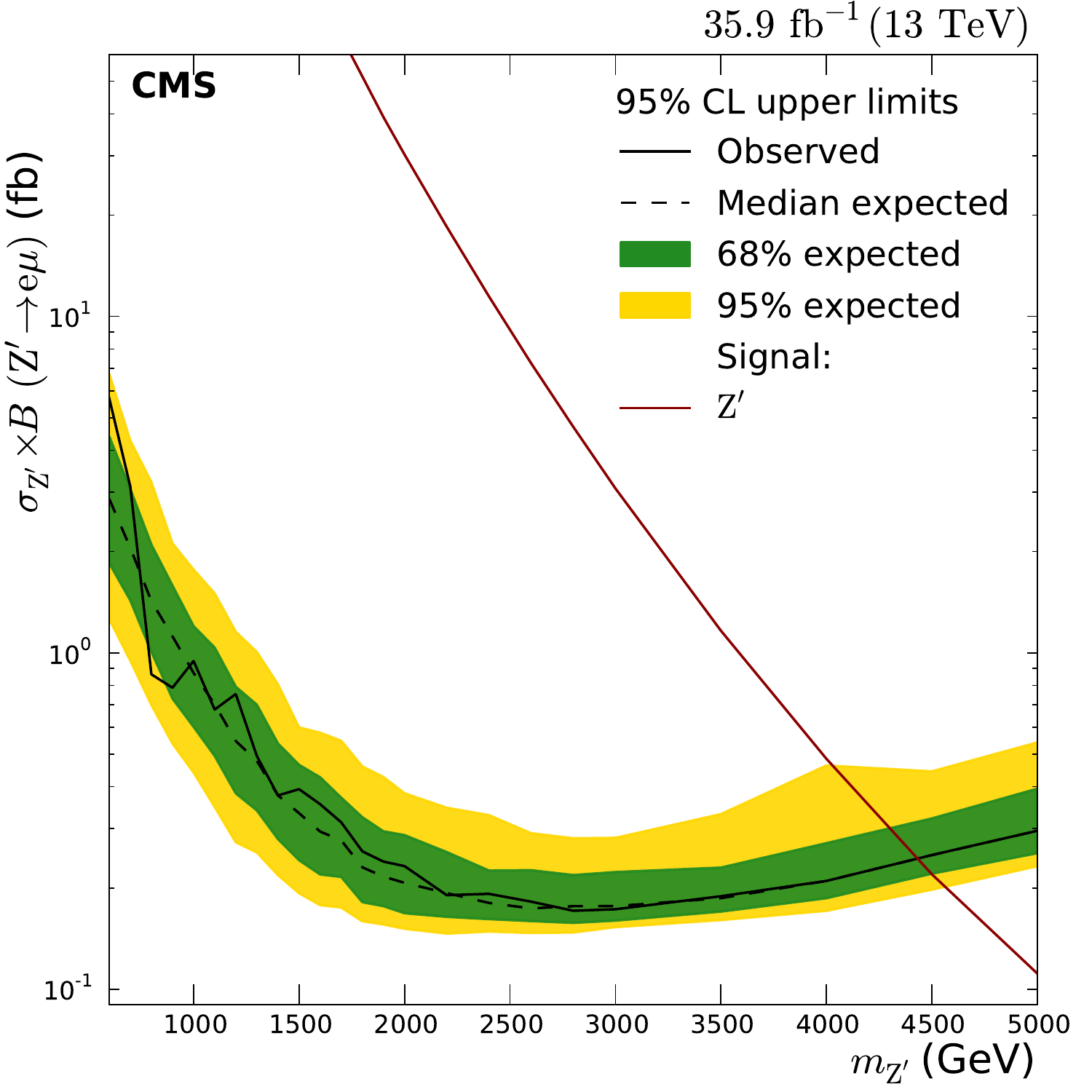}
\caption{The upper limits at 95\% CL on the product of the signal cross section and the branching fraction, assuming $\mathcal{B} = 10\%$
         for the decay $\PZpr\to\Pe\Pgm$, as a function of $m_{\PZpr}$. The 68 and 95\% CL
         intervals on the median are indicated, respectively, by the inner green and outer yellow shadings.
}
\label{fig:ZPrimelimit}
\end{figure}

In the QBH search, we set mass limits on the production threshold in models with $n = 1$ (RS) and $n>1$ (ADD)
extra dimensions. The limits at 95\% CL on $\sigma \mathcal{B}$ for the QBH signal are shown in Fig.~\ref{fig:QBHlimit}.
The observed and expected lower mass limits on $m_\text{th}$ are numerically the same and correspond to 3.6,
5.3, 5.5, and 5.6\TeV for $n = 1$, 4, 5, and 6, respectively. In the ADD model, the results are
given for $n = 4$, 5, and 6.
The expected and observed limits at 95\% CL on the \PZpr mass are also the same and equal to 4.4\TeV, as shown in
Fig.~\ref{fig:ZPrimelimit}.
With increasing \PZpr boson mass, the phase space for the on-shell \PZpr production in $\Pp\Pp$ collisions at 13\TeV decreases
because of decreasing parton-parton luminosity, leading to the production of an increasing fraction of off-shell objects at lower masses.
This effect leads to weaker \PZpr boson mass limits at higher mass values.
The observed limit looks smoother than for the RPV signal (Fig.~\ref{fig:ZPrimelimit}), because there are fewer signal mass hypotheses probed. The results
presented in this paper also put constraints on specific models involving LFV \PZpr such as proposed in Refs.~\cite{Frere:2015tea}.

\clearpage
\section{Summary}
A search for heavy resonances decaying into $\Pe\Pgm$ pairs has been carried out in proton-proton collisions, recorded with the CMS detector at the LHC
at a center-of-mass energy of 13\TeV, corresponding to an integrated luminosity of 35.9\fbinv.
Good agreement is observed between the data and the standard model expectation.
Limits are set on the resonant production of {\Pgt} sneutrinos ($\PSgn_\Pgt$) in
R-parity violating supersymmetric models.
For couplings $\lambda_{132} = \lambda_{231} = \lambda'_{311} = 0.01$ and 0.1, a $\PSgn_\Pgt$ is excluded for masses below 1.7 and 3.8\TeV respectively,
assuming it is the lightest supersymmetric particle.
Lower limits of 5.3, 5.5, and 5.6\TeV are set on the threshold mass of quantum black holes in a model with 4, 5, and 6
large extra spatial dimensions, respectively.
For the model with a single, warped extra spatial dimension, the lower limit on the threshold mass is 3.6\TeV.
Also, a \PZpr boson with a 10\% branching fraction to the $\Pe\Pgm$
channel is excluded for masses below 4.4\TeV.
In all cases, the results of this search
improve the previous lower limits by about 1\TeV.

\begin{acknowledgments}
We congratulate our colleagues in the CERN accelerator departments for the excellent performance of the LHC and thank the technical and administrative staffs at CERN and at other CMS institutes for their contributions to the success of the CMS effort. In addition, we gratefully acknowledge the computing centers and personnel of the Worldwide LHC Computing Grid for delivering so effectively the computing infrastructure essential to our analyses. Finally, we acknowledge the enduring support for the construction and operation of the LHC and the CMS detector provided by the following funding agencies: BMWFW and FWF (Austria); FNRS and FWO (Belgium); CNPq, CAPES, FAPERJ, and FAPESP (Brazil); MES (Bulgaria); CERN; CAS, MoST, and NSFC (China); COLCIENCIAS (Colombia); MSES and CSF (Croatia); RPF (Cyprus); SENESCYT (Ecuador); MoER, ERC IUT, and ERDF (Estonia); Academy of Finland, MEC, and HIP (Finland); CEA and CNRS/IN2P3 (France); BMBF, DFG, and HGF (Germany); GSRT (Greece); OTKA and NIH (Hungary); DAE and DST (India); IPM (Iran); SFI (Ireland); INFN (Italy); MSIP and NRF (Republic of Korea); LAS (Lithuania); MOE and UM (Malaysia); BUAP, CINVESTAV, CONACYT, LNS, SEP, and UASLP-FAI (Mexico); MBIE (New Zealand); PAEC (Pakistan); MSHE and NSC (Poland); FCT (Portugal); JINR (Dubna); MON, RosAtom, RAS, RFBR and RAEP (Russia); MESTD (Serbia); SEIDI, CPAN, PCTI and FEDER (Spain); Swiss Funding Agencies (Switzerland); MST (Taipei); ThEPCenter, IPST, STAR, and NSTDA (Thailand); TUBITAK and TAEK (Turkey); NASU and SFFR (Ukraine); STFC (United Kingdom); DOE and NSF (USA).

\hyphenation{Rachada-pisek} Individuals have received support from the Marie-Curie program and the European Research Council and Horizon 2020 Grant, contract No. 675440 (European Union); the Leventis Foundation; the A. P. Sloan Foundation; the Alexander von Humboldt Foundation; the Belgian Federal Science Policy Office; the Fonds pour la Formation \`a la Recherche dans l'Industrie et dans l'Agriculture (FRIA-Belgium); the Agentschap voor Innovatie door Wetenschap en Technologie (IWT-Belgium); the F.R.S.-FNRS and FWO (Belgium) under the ``Excellence of Science - EOS'' - be.h project n. 30820817; the Ministry of Education, Youth and Sports (MEYS) of the Czech Republic; the Council of Science and Industrial Research, India; the HOMING PLUS program of the Foundation for Polish Science, cofinanced from European Union, Regional Development Fund, the Mobility Plus program of the Ministry of Science and Higher Education, the National Science Center (Poland), contracts Harmonia 2014/14/M/ST2/00428, Opus 2014/13/B/ST2/02543, 2014/15/B/ST2/03998, and 2015/19/B/ST2/02861, Sonata-bis 2012/07/E/ST2/01406; the National Priorities Research Program by Qatar National Research Fund; the Programa Severo Ochoa del Principado de Asturias; the Thalis and Aristeia programmes cofinanced by EU-ESF and the Greek NSRF; the Rachadapisek Sompot Fund for Postdoctoral Fellowship, Chulalongkorn University and the Chulalongkorn Academic into Its 2nd Century Project Advancement Project (Thailand); the Welch Foundation, contract C-1845; and the Weston Havens Foundation (USA).
\end{acknowledgments}

\bibliography{auto_generated}

\clearpage
\appendix
\section{Additional analysis plots}

\begin{figure}[h]
\centering
\includegraphics[width=0.90\textwidth]{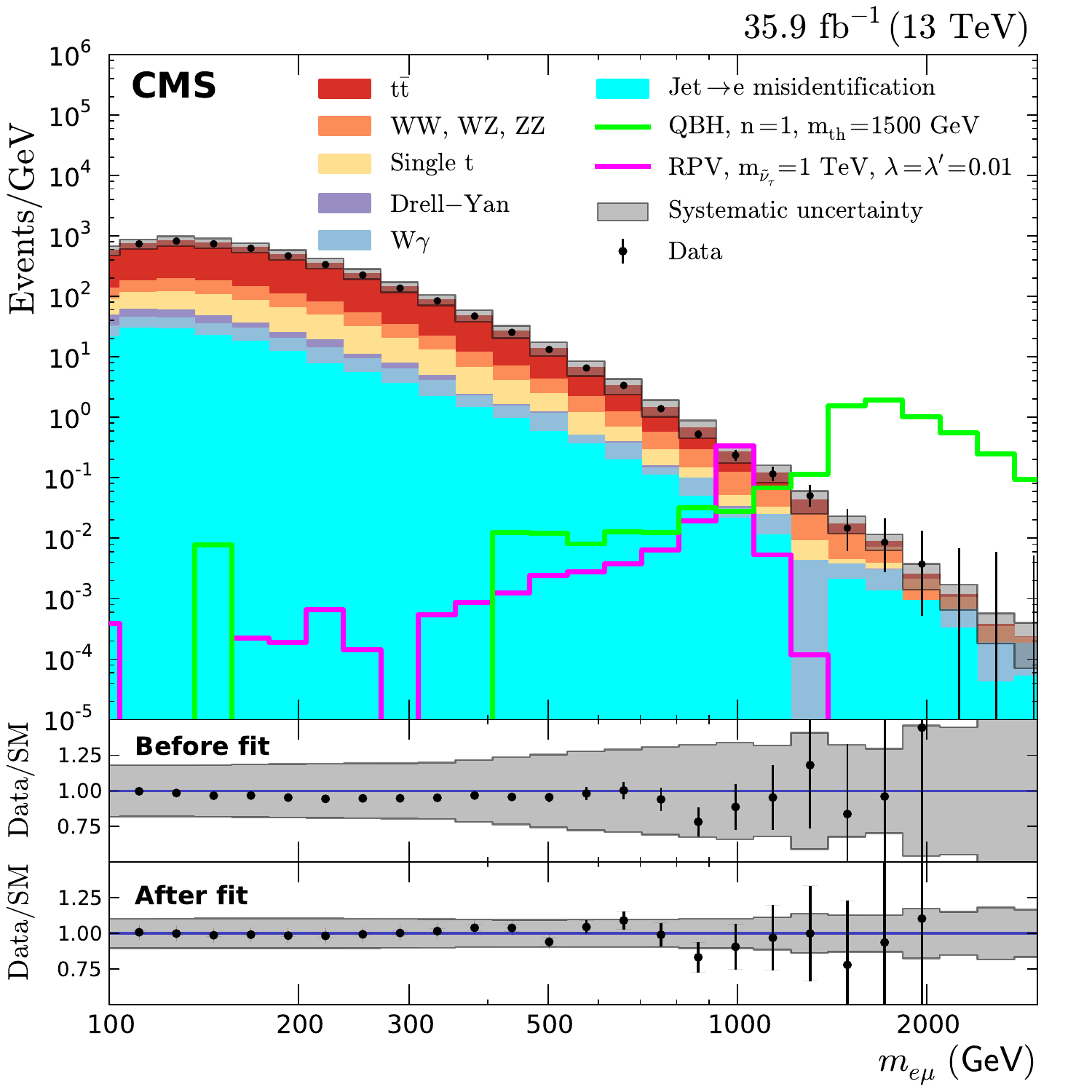}
\captionof{figure}[]{Invariant mass of the {\Pe}{\Pgm} pair for all events that pass the event selection criteria. In the lower panels, we show
the ratio of the data to the before-fit and after-fit background predictions, including uncertainties. The label $\lambda$ stands for
$\lambda_{132} = \lambda_{231}$, while $\lambda'$ stands for $\lambda'_{311}$. The content is the same as in Fig.~\ref{fig:emumass},
but with a coarser binning.}
\label{fig:emumass_simple}
\end{figure}

\begin{figure}[h]
\centering
\includegraphics[width=0.75\textwidth]{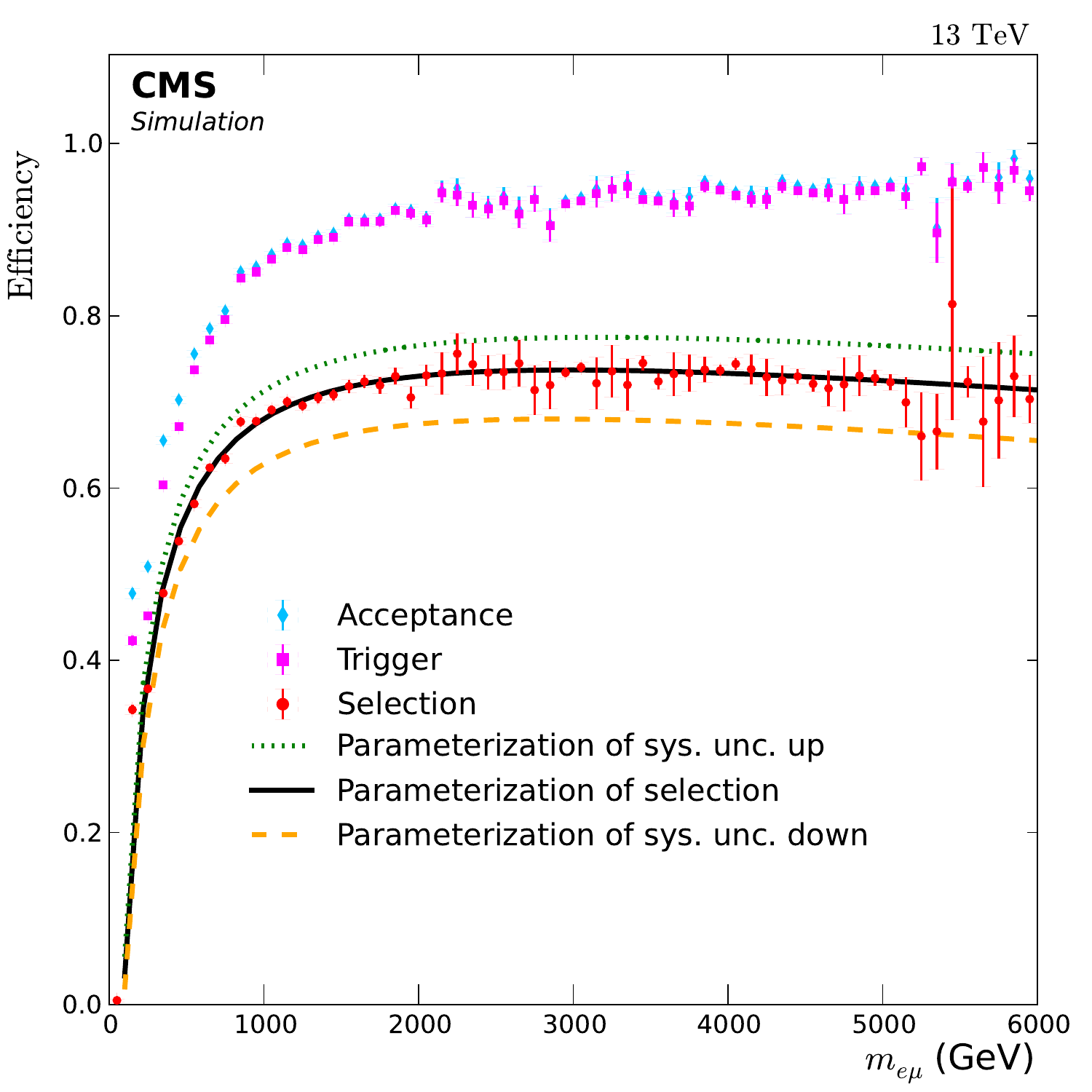}
\captionof{figure}[]{Efficiency of the RPV signal for all events after the acceptance requirements (light blue triangular points), after
                     acceptance and trigger requirements (magenta square points), and after the full selection, which includes acceptance and
                     trigger criteria (red round points). The reconstruction efficiency is also included, with the product of the final acceptance
                     and efficiency parametrized for the statistical interpretation by a function illustrated by the black line. The
                     systematic uncertainties are obtained by propagating the effect of the systematic uncertainties to the efficiency.
                     The systematically shifted upper and lower efficiency points are not shown in the figure, but just the parametrization
                     of both dependencies, with upward shifts in dotted green and downward shifts in dashed orange.}
\label{fig:emu_eff}
\end{figure}

\begin{figure}[h]
\centering
\includegraphics[width=0.65\textwidth]{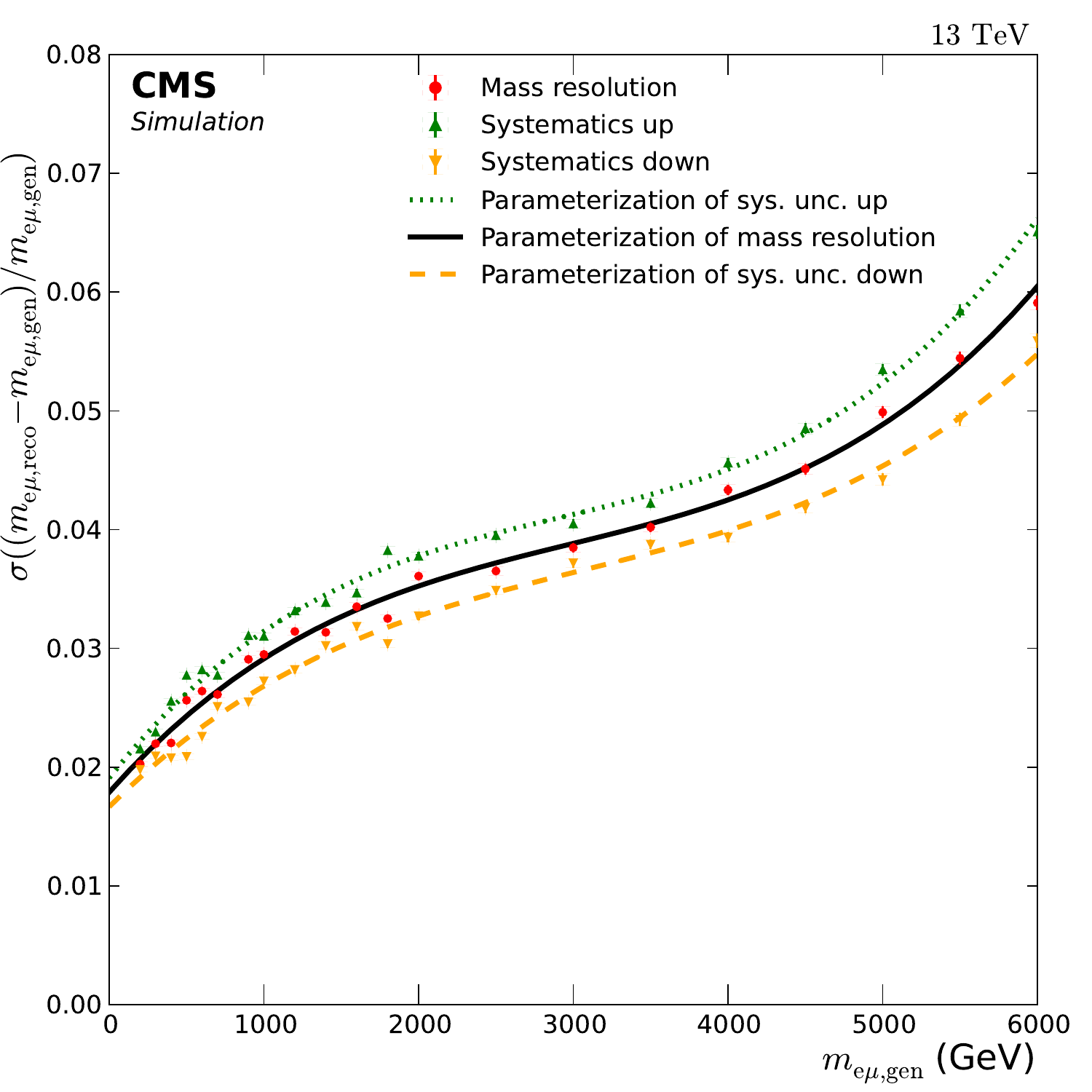}
\captionof{figure}[]{Relative mass resolution for all {\Pe}{\Pgm} pairs, obtained through simulation of the RPV signal, from the reconstructed mass
                     $m_{\Pe\Pgm,\text{reco}}$ and the generated mass $m_{\Pe\Pgm,\text{gen}}$, as a function of the generated
                     mass. The effect of the systematic uncertainties on the mass resolution is shown. The systematically shifted upper and
                     lower mass resolutions are shown in the figure with the corresponding parametrization for the upward shifts in dotted green
                     and the downward shifts in dashed orange.}
\label{fig:emu_res}
\end{figure}

\clearpage

\cleardoublepage \section{The CMS Collaboration \label{app:collab}}\begin{sloppypar}\hyphenpenalty=5000\widowpenalty=500\clubpenalty=5000\textbf{Yerevan Physics Institute,  Yerevan,  Armenia}\\*[0pt]
A.M.~Sirunyan, A.~Tumasyan
\vskip\cmsinstskip
\textbf{Institut f\"{u}r Hochenergiephysik,  Wien,  Austria}\\*[0pt]
W.~Adam, F.~Ambrogi, E.~Asilar, T.~Bergauer, J.~Brandstetter, E.~Brondolin, M.~Dragicevic, J.~Er\"{o}, A.~Escalante Del Valle, M.~Flechl, M.~Friedl, R.~Fr\"{u}hwirth\cmsAuthorMark{1}, V.M.~Ghete, J.~Grossmann, J.~Hrubec, M.~Jeitler\cmsAuthorMark{1}, A.~K\"{o}nig, N.~Krammer, I.~Kr\"{a}tschmer, D.~Liko, T.~Madlener, I.~Mikulec, E.~Pree, N.~Rad, H.~Rohringer, J.~Schieck\cmsAuthorMark{1}, R.~Sch\"{o}fbeck, M.~Spanring, D.~Spitzbart, A.~Taurok, W.~Waltenberger, J.~Wittmann, C.-E.~Wulz\cmsAuthorMark{1}, M.~Zarucki
% ...
\vskip\cmsinstskip
% ...
1:~~Also at Vienna University of Technology, Vienna, Austria\\
% ...

\end{sloppypar}
\end{document}